\documentclass[conference]{IEEEtran}

\usepackage{cite}
\usepackage{amsmath,amssymb,amsfonts}
\usepackage{iftex}
\ifPDFTeX
  \usepackage[T1]{fontenc}
  \usepackage[utf8]{inputenc}
  
\else
  \usepackage{fontspec}
\fi
\usepackage{graphicx}
\usepackage[table]{xcolor}
\usepackage{url}
\usepackage{microtype}

\graphicspath{{figures/}}
\usepackage{makecell}
\usepackage{tabularx}
\usepackage{enumitem}
\usepackage[font=footnotesize]{caption}
\usepackage{tikz}
\usepackage{booktabs}
\usepackage{multirow}
\usepackage{pgfplots}
\usepackage{subcaption}

\usepackage{siunitx}
\usepackage{algorithm}
\usepackage{algpseudocode}
\usepackage{listings}
\definecolor{apilightblue}{RGB}{70, 150, 220}
\usepackage{pifont}
\usepackage[normalem]{ulem}
\usepackage[hidelinks]{hyperref}

\pgfplotsset{compat=1.18}
\newif\ifrevision
\revisiontrue

\ifrevision
  \long\def\del#1{{\color{red}#1}}    
  \long\def\delx#1{{\color{red}#1}}   
  \newcommand{\note}[1]{\textcolor{violet}{\textbf{[TODO: #1]}}}  
\else
  
  \newcommand{\del}[1]{}
  \newcommand{\delx}[1]{}
  \newcommand{\note}[1]{}
\fi

\lstdefinestyle{vectraflow}{
  language=Python,
  basicstyle=\ttfamily\small,
  keywordstyle=\bfseries\color{blue},
  commentstyle=\itshape\color{gray},
  stringstyle=\color{teal},
  showstringspaces=false,
  breaklines=true,
  frame=lines,
  numbers = left,
  tabsize=2
}

\begin{document}

\title{Continuous Prompts: Adaptive Semantic Query Processing over Unstructured Streams}

\author{
\IEEEauthorblockN{Shu Chen}
\IEEEauthorblockA{Brown University\\
Providence, USA\\
shu\_chen@brown.edu}
\and
\IEEEauthorblockN{Deepti Raghavan}
\IEEEauthorblockA{Brown University\\
Providence, USA\\
deeptir@brown.edu}
\and
\IEEEauthorblockN{U\u{g}ur \c{C}etintemel}
\IEEEauthorblockA{Brown University\\
Providence, USA\\
ugur\_cetintemel@brown.edu}
}

\maketitle

\begin{abstract}
Continuous analytics on unstructured streams needs stateful semantic queries whose execution adapts to shifting arrival rates and content, balancing throughput and answer quality under fixed compute resources. We present \textit{Continuous Prompts} (CPs), a framework for unbounded streams built on three streaming-native constructs: semantic windows, dynamic semantic group-by, and continuous RAG. These segment streams by content, track evolving semantic groups, and match new data to persistent retrieval intents. We build and evaluate implementations based on generative LLMs, decision models, and embeddings, revealing complementary throughput--quality trade-offs.

CPs jointly optimize operator implementations, model selection, tuple batching, and operator fusion. All these choices affect throughput and answer quality,  making their workload-dependent quality effects explicit planning considerations. Our planner uses cost-aware \textit{multi-objective Bayesian optimization} (MOBO) to learn pipeline throughput--quality frontiers under limited profiling budgets. Operator-level probes refine estimates across candidate plans, reducing the need for full-pipeline profiling. At runtime, CPs select the predicted highest-quality plan that keeps pace with arrivals, while shadow evaluation detects drift and triggers targeted re-profiling.

We implement CPs in VectraFlow and evaluate operator microbenchmarks and two end-to-end monitoring pipelines, including comparisons with a per-tuple LLM baseline executed in Apache Flink. On the same hardware, CPs achieve up to $9.5\times$ the throughput of this baseline within 4\% of its accuracy. Further experiments demonstrate efficient frontier discovery, quality-aware plan selection across arrival rates, and targeted frontier recovery under distribution shifts.




\end{abstract}


\section{Introduction}

Applications such as financial surveillance, clinical monitoring, and
regulatory compliance require continuous semantic analysis of evolving
unstructured data. Consider a financial news monitor that tracks a
portfolio and continuously surfaces consequential developments. It must
distinguish an earnings warning from routine commentary, even when both
are highly similar to a company's profile, and organize incoming reports
as events emerge and evolve. These tasks require interpreting content
against user-defined logic and maintaining semantic state across
arrivals. They also impose an execution challenge: under fixed compute
resources, the system must balance answer quality against the throughput
needed to keep pace with evolving workloads.

Existing systems address parts of this problem. Palimpzest~\cite{liu2025palimpzest,russo2025abacus},
Lotus~\cite{patel2025semanticoperators}, and
DocETL~\cite{shankar2025docetl} provide semantic operators and optimization
for LLM-based analytics over static datasets, but do not address
persistent execution over unbounded, evolving streams. General-purpose
stream processors such as Apache Flink support per-tuple model inference
through \texttt{ML\_PREDICT}~\cite{flink22}, but treat these invocations as
opaque model calls, without semantic operators or accuracy-aware
optimization. VectraFlow~\cite{Lu2025VectraFlow} introduced continuous
embedding-based operators, including vector filters and vector top-$k$.
These operators support efficient similarity-based processing, but
monitoring queries also require judgments about significance,
relationships, and event continuity that vector similarity alone does
not directly express. Together, these limitations motivate a framework
that combines stateful semantic operators with quality-aware adaptive
execution.

We introduce \textit{Continuous Prompts} (CPs), a framework that lifts
one-shot prompt execution into persistent semantic computations over
unstructured streams. Just as continuous
queries~\cite{terry1992continuous} extend relational query processing
to unbounded data, CPs extend semantic query processing with operators
that maintain evolving semantic state. CPs introduce three streaming-native
constructs. \textit{Semantic windows} segment streams according to
content-dependent continuity rather than fixed time or tuple boundaries.
\textit{Dynamic semantic group-by} forms and updates meaning-based
groups as new content arrives. \textit{Continuous RAG} matches incoming
stream items against persistent retrieval intents. Each construct supports alternative LLM-, decision-model-, and embedding-based implementations, offering complementary throughput–quality trade-offs.

Efficient execution requires choosing among these implementations and
models, as well as deciding how to batch tuples and fuse adjacent
operators. Model and implementation selection offer familiar ways to
trade quality for throughput in semantic processing
systems~\cite{liu2025palimpzest,patel2025semanticoperators,shankar2025docetl,Lu2025VectraFlow}.
Tuple batching and operator fusion introduce another dimension:
they change how the same model is invoked. Batching amortizes
instructions and invocation overhead across tuples, while fusion avoids
repeated processing of shared inputs across operators. Unlike their
classical counterparts, however, these transformations need not preserve
semantic answers. Tuples in a batch can interfere, and combined operator instructions can alter outputs. Quality impacts depend on the operator, model, and input data, so they require explicit planning. A plan that works for one workload may fail as arrival rates increase or content drifts.

We address this challenge with a dynamic planning framework that learns
and maintains throughput--quality Pareto frontiers over the combined
plan space. Our planner uses cost-aware \textit{multi-objective Bayesian
optimization} (MOBO) to allocate a limited profiling budget to informative
operator-level probes. Each probe refines estimates for multiple
candidate plans, reducing the need for full-pipeline profiling. At
runtime, the planner uses these estimates to select a high-quality plan
that can sustain the current arrival rate. Shadow executions monitor
operator behavior, detect distribution shifts, and trigger targeted
re-profiling of affected operators. This approach decouples inexpensive plan
selection as load changes from frontier updates when the underlying
performance and quality trade-offs change.

We implement CPs in VectraFlow and evaluate operator-level
microbenchmarks and two end-to-end monitoring pipelines over real data,
including comparisons with per-tuple LLM execution in Apache Flink.
On the same hardware, CPs achieve up to $9.5\times$ the throughput of
this baseline within $4\%$ of its accuracy, and over $11\times$ when
larger quality reductions are acceptable. Further experiments evaluate
frontier discovery under limited profiling budgets, plan selection
across arrival rates, and targeted frontier recovery under distribution
shifts.

In summary, this paper makes three contributions:
\begin{itemize}
    \item We introduce Continuous Prompts (CPs) and define new
    streaming-native semantic constructs: semantic windows, dynamic
    semantic group-by, and continuous RAG. We also describe their implementations based
    on generative LLMs, decision models, and embeddings
    (Section~\ref{sec:streaming-ops}).

    \item We develop a dynamic planning framework that uses cost-aware
    MOBO to learn throughput--quality Pareto frontiers over operator
    implementations, model selection, tuple batching, and operator
    fusion, and supports runtime plan selection and targeted
    re-profiling under limited probing budgets and drift (Section~\ref{sec:dynamic-planning}).

    \item We implement CPs in VectraFlow and evaluate them on operator-level microbenchmarks and end-to-end streaming pipelines over real data (Sections~\ref{sec:streaming-ops} and~\ref{sec:evaluation}).
\end{itemize}

The rest of the paper is organized as follows. Section~\ref{sec:vectraflow} introduces the VectraFlow architecture and system model. Section~\ref{sec:streaming-ops} presents the streaming-native semantic operators and their microbenchmarks. Section~\ref{sec:dynamic-planning} describes the dynamic planning framework, including the plan space, MOBO-based Pareto frontier learning, and online adaptation. Section~\ref{sec:evaluation} presents the experimental evaluation. Sections~\ref{sec:related} and~\ref{sec:conclusion} discuss related work and conclude.

\section{System Architecture}
\label{sec:vectraflow}
The Continuous Prompts framework is built on the VectraFlow stream processing system~\cite{Lu2025VectraFlow}. We extend VectraFlow with continuous semantic operators, multiple operator implementations, and adaptive optimization mechanisms for persistent LLM-native pipelines.

\noindent\textbf{Streaming Model.}
The system ingests unbounded streams of unstructured content such as clinical notes, financial news, or social media posts, arriving at a time-varying rate $\lambda(t)$.
These inputs are heterogeneous in length, style, and vocabulary, often lacking a predefined schema or fixed attributes.

\noindent\textbf{Data Model.}
Incoming data is modeled as timestamped tuples $\mathcal{S} = \{ (t_i, x_i) \mid i = 1, 2, \ldots \}$ under an extended relational model, where $t_i$ is the arrival time and $x_i$ is the payload containing one or more unstructured attributes consumed by semantic operators. We assume
tuples arrive in timestamp order. Out-of-order
arrival can be handled by the watermarking of the host
engine~\cite{carbone2015flink} and is orthogonal to this work.

\noindent\textbf{Query Model.}
A query plan is a directed acyclic graph of \emph{semantic operators} $\mathcal{O}=\{o_1,\dots,o_k\}$ that filter, map, segment, group, or rank tuples (Table~\ref{tab:semantic-ops}). Each operator admits several implementations, realized by a generative LLM, a decision model, or an embedding function. All operators except \texttt{sem\_filter} and \texttt{sem\_map} are stateful: they maintain bounded internal state updated per tuple, which lets them adapt to semantic drift in the stream. \texttt{sem\_agg} and \texttt{sem\_topk} also support incremental execution within a window.

\noindent\textbf{Execution Model.}
Operators execute in a pipelined, non-blocking fashion: each consumes tuples from its input queue as they arrive, so operators placed on different resources process different tuples concurrently. The deployment assigns each operator to a model server; LLM calls are served by vLLM~\cite{kwon2023pagedattention}, which schedules up to $P$ concurrent requests per server with continuous batching~\cite{yu2022orca}.

\noindent\textbf{Cost Model.}
Each operator $o_j$ carries a performance profile $C(o_j) = \langle \text{throughput}(o_j),\, \text{accuracy}(o_j) \rangle$ that drives runtime choices of operator implementation, model, batch size, and fusion strategy (Section~\ref{sec:dynamic-planning}).

\begin{table*}[t]
\centering
\setlength{\tabcolsep}{5pt}
\renewcommand{\arraystretch}{1.15}
\small
\begin{tabular}{@{}p{2.0cm} p{1.9cm} p{1.15cm} p{5.6cm} p{5.45cm}@{}}
\toprule
\textbf{Operator} & \textbf{Signature} & \textbf{State} & \textbf{Description} & \textbf{Example Use Case} \\
\midrule
\texttt{sem\_filter}
& $S \rightarrow S$ & --
& Keeps tuples that satisfy a semantic predicate over topic, sentiment, or entity.
& Select tweets related to \emph{Ukraine} or \emph{COVID-19} from a mixed news stream. \\
\texttt{sem\_map}
& $S \rightarrow S$ & --
& Transforms unstructured text into structured records such as entities, relations, or JSON key--value pairs.
& Extract company name and event from news headlines. \\
\texttt{sem\_agg}
& $\langle R \rangle \rightarrow S$ & $O(1)$
& Summarizes each window incrementally via \texttt{init()}, \texttt{increment()}, \texttt{finalize()}.
& Continuously summarize market sentiment over the last 100 tweets mentioning Apple (\$AAPL). \\
\texttt{sem\_topk}
& $\langle R \rangle \rightarrow S$ & $O(k)$
& Scores each tuple on arrival against a criterion (e.g., impact, novelty) and keeps the $k$ highest-scoring tuples per window.
& Select the 10 most influential financial news items as the market evolves. \\
\texttt{sem\_join}
& $S \times S \rightarrow S$ & $O(W)$
& Correlates tuples across streams based on semantic similarity rather than strict key equality.
& Match analyst reports to relevant stock price movements. \\
\midrule
\mbox{\texttt{sem\_window}\,$\star$}
& $S \rightarrow \langle R \rangle$ & $O(W_{\max})$
& Closes a window when the arriving tuple's continuity with the open window falls below a threshold.
& Detect when discourse shifts from ``peace talks'' to ``sanctions'' in the Ukraine event stream. \\
\mbox{\texttt{sem\_groupby}\,$\star$}
& $S \rightarrow S$ & $O(G_{\max})$
& Routes each tuple to one group in a catalog that is either declared up front (static) or evolves as groups emerge, merge, and retire (dynamic).
& Track evolving news events and continuously update group assignments as underlying topics shift. \\
\mbox{\texttt{cts\_filter}\,$\star$}
& $S \times \mathcal{P} \rightarrow S$ & $O(|\mathcal{P}|)$
& Emits tuples relevant to at least one prompt in a persistent store derived from a user reference table.
& Continuously surface financial news relevant to a given stock portfolio. \\
\mbox{\texttt{cts\_topk}\,$\star$}
& \mbox{$\langle R \rangle \times \mathcal{P} \rightarrow S$} & $O(k|\mathcal{P}|)$
& Retains the $k$ tuples per window ranked highest against each stored prompt, maintaining one ranked list per prompt.
& Surface the 3 most material news items per holding in each window. \\
\bottomrule
\end{tabular}
\caption{Overview of semantic operators in CPs. $S$ denotes a
stream, $\langle R \rangle$ a sequence of windows, and $\mathcal{P}$ a
persistent prompt store; state is the operator's working state,
bounded in every case rather than growing with stream length. Window
operators are the only source of $\langle R \rangle$, so the blocking
operators that consume it require an upstream window. Operators marked
with $\star$ are streaming-native and are formally defined in
Section~\ref{sec:streaming-ops}; the remainder follow prior semantic
operator systems~\cite{patel2025semanticoperators, liu2025palimpzest}.}
\label{tab:semantic-ops}
\end{table*}

\section{Streaming-Native Semantic Operators}
\label{sec:streaming-ops}

We introduce three streaming-native constructs for adaptive processing over unstructured data streams: semantic windows, dynamic semantic group-by, and continuous RAG, which provides two operators, a continuous filter and a continuous top-$k$. For each, we present LLM-based, Jev-based, and embedding-based implementations, whose APIs are shown in Figure~\ref{fig:streaming-ops-api}, and analyze their behavior through microbenchmarks.

\begin{figure}[t]
\begin{lstlisting}[
  language=Python,
  basicstyle=\scriptsize\ttfamily,
  stringstyle=\color{teal},
  commentstyle=\color{gray}\itshape,
  keywordstyle=\color{violet},
  morekeywords={sem_window, sem_groupby, cts_filter, cts_topk},
  numbers=left,
  numberstyle=\tiny\color{gray},
  numbersep=4pt,
  xleftmargin=1.2em,
  framexleftmargin=1.2em,
  frame=lines,
  rulecolor=\color{black},
  columns=fixed,
  keepspaces=true,
  breaklines=false,
  showstringspaces=false,
  aboveskip=4pt,
  belowskip=4pt]
# sem_window: "llm" (M1; strategy="summary" for M2),
#   "embedding" (M3), or "jev" (M4)
window = sem_window(method="llm", strategy="pairwise",
  instruction="Rate from 0 to 1 how likely the second"
    " item continues the same event as the first.")
# sem_groupby: "llm" (M1/M2), "embedding" (M3/M4), or
#   "jev" (M5); refine=True selects M2/M4
groupby = sem_groupby(method="embedding", refine=True,
  text="news headline")
# cts_filter / cts_topk: "llm", "jev", "embedding",
#   or a cascade (X, Y)
cts = cts_filter(reference=stock_portfolio,
  method=("embedding", "jev"),
  intent="Find news that would affect"
    " my {stock_portfolio}.")
top = cts_topk(reference=stock_portfolio, k=3,
  method=("embedding", "llm"),
  intent="Surface news most likely to move"
    " my {stock_portfolio}.")
\end{lstlisting}
\caption{API usage examples for the streaming-native operators. Each
 operator selects its implementation through \texttt{method}; for
 \texttt{cts\_filter} and \texttt{cts\_topk}, a pair
 (\texttt{X}, \texttt{Y}) denotes a cascade in which \texttt{X}
 shortlists the top prompt instances and \texttt{Y}
 verifies or scores them.}
\label{fig:streaming-ops-api}
\end{figure}

\subsection{Semantic Windows}
Windows are the standard mechanism for imposing finite scopes on an
unbounded stream. In the stream-to-relation view~\cite{cql}, a window
operator maps a stream $S$ to a sequence of finite relations
$\langle R_1, R_2, \dots\rangle$ that downstream operators consume.
 Tumbling and sliding windows cut the stream at fixed offsets in time or tuple count, and session windows~\cite{dataflow}
close after a gap in arrivals. Frames~\cite{frames} go one
step further and use attribute values, closing a window when a
condition over a structured attribute holds, such as a value staying
above a threshold for a minimum number of tuples. Outside stream
engines, topic segmentation~\cite{hearst1997texttiling} and first story
detection~\cite{allan1998tdt,petrovic2010fsd} do detect content shifts
in text, but with a fixed lexical or vector similarity rather than a
user-specified criterion.

However, none of these windows can place a boundary based on the
content of an unstructured attribute. When election coverage moves from
``campaign'' to ``results'' articles, time- and count-based windows
cut regardless of the shift, a session window closes only when
articles stop arriving, and a
frame would need a structured attribute marking the transition.

\noindent\textbf{Definition.}
We therefore propose \emph{semantic windows}, which evaluate a
continuity function over the unstructured content of each tuple,
detecting signals such as topic shifts, sentiment changes, or new
entity mentions. Like its classical counterparts, \texttt{sem\_window} only assigns
tuples to windows; it computes nothing over them. It plays the role of
\texttt{AssignWindows} in the Dataflow model~\cite{dataflow}, marking
where one window ends and the next begins, and is always paired with a
downstream operator that consumes the windows it produces. In the
example above, \texttt{sem\_window} delimits each storyline and a
downstream \texttt{sem\_agg} or \texttt{sem\_topk} computes the summary
or the most impactful items within it.

The operator maintains a single active context $\kappa_t$ that represents
the currently open window $W$, such as its most recent tuple, a running
summary, or an embedding centroid, depending on the implementation. It closes $W$ at the first tuple whose
continuity with the context falls below a threshold $\tau$. As with a
session window under continuous arrivals, a window may never close if
content drifts slowly enough that no tuple crosses $\tau$, so we also
cap the window at a maximum length $W_{\max}$. A window therefore closes at $x_t$ when
\[
\mathrm{cont}(x_t, \kappa_{t-1}) < \tau \quad \lor \quad |W| \ge W_{\max},
\]
where the first condition is the semantic boundary and the second guarantees closure; $x_t$ then opens the next window. Threshold calibration is discussed in Section~\ref{par:threshold-calibration}.

\noindent\textbf{Implementation.}
 We implement four strategies that differ in how the active context
$\kappa_t$ is represented and how continuity is judged, trading
boundary accuracy against throughput; interfaces are shown in
 Figure~\ref{fig:streaming-ops-api}.
\emph{M1 (Pairwise)} represents the context by the most recent tuple
and computes $\mathrm{cont}(x_t, x_{t-1})$ between consecutive tuples
with one LLM call, opening a new window when the score drops below
$\tau$.
\emph{M2 (Summary-based)} represents the context by a single evolving summary of the open window. Rather than scoring $x_t$ and updating the summary in two
calls, a single call returns the continuity decision
together with the extended summary or
a fresh summary of $x_t$, halving the invocations per tuple.
Comparing against the whole window rather than only the previous tuple
tolerates local fluctuation: a tuple that differs slightly from its
predecessor but stays on the window's topic does not open a new window.
\emph{M3 (Embedding-based)} represents the context by a running
centroid over the embeddings of the open window; $x_t$ extends the
current window if cosine similarity to the centroid is $\geq \tau$;
otherwise, a boundary is placed and the centroid is reinitialized.
The centroid is updated as an incremental mean, so
no past embeddings are kept.
\emph{M4 (Jev-based)} keeps M1's pairwise context but replaces the
generative LLM with a Jev decision model~\cite{typesafejev}, which takes
$(x_{t-1}, x_t)$ and returns a yes/no decision on whether $x_t$
continues the current window. Because the model emits a decision
rather than generating text, each tuple incurs one lightweight
invocation.

\begin{figure*}[t]
 \centering
 \includegraphics[width=\textwidth]{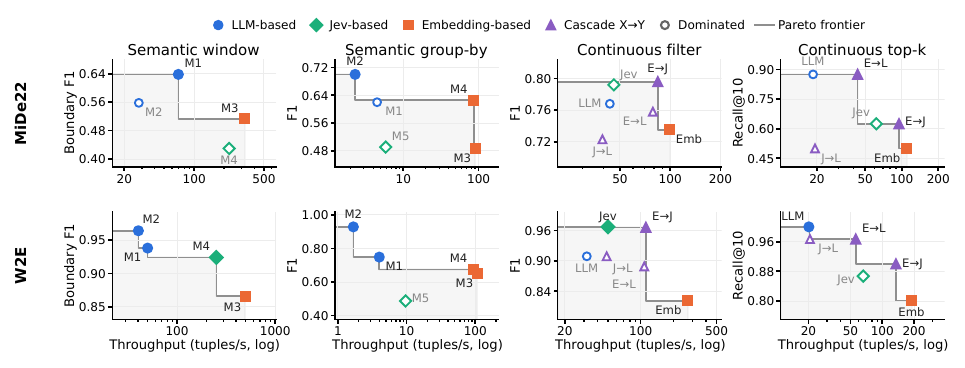}
 \caption{Throughput--accuracy trade-offs of the streaming-native
 operators on MiDe22 (top) and W2E (bottom). Each column is one
 operator, and the shaded region is dominated by the Pareto frontier.}
 \label{fig:micro-tradeoffs}
\end{figure*}

\noindent\textbf{Evaluation.}
We evaluate the four strategies on two event streams:
MiDe22~\cite{toraman2024mide22}, which consists of temporally ordered
tweets spanning 40 real-world events, and W2E~\cite{hoang2018w2e}, a large-scale news event stream. Figure~\ref{fig:micro-tradeoffs}
collects the trade-offs for every operator in this section.
Unless otherwise noted, all LLM-based operators in this section use
Qwen2.5-7B-Instruct~\cite{yang2025qwen} served by
vLLM~\cite{kwon2023pagedattention} on a single GPU at the maximum
sustainable request concurrency; all Jev-based operators use
Open-Jev-2B~\cite{openjev}, an open-weight decision model; and all
embedding-based operators use all-MiniLM-L6-v2~\cite{wang2020minilm}. Throughout the paper, we
measure accuracy against the dataset's labels whenever the task has
them; for tasks without labels, such as impact ranking or free-form
extraction, we follow prior semantic operator
systems~\cite{patel2025semanticoperators,liu2025palimpzest} and measure
agreement with a gold model, GPT-4o~\cite{openai2024gpt4o}, run on the
same inputs with the same instructions. Since windowing is a
segmentation task, we report boundary F1, counting a predicted boundary
as correct only if it matches the ground-truth event transition in
the dataset labels. We also evaluated tolerances of one and two tuples; they
leave the ranking unchanged, so we omit them. We set $W_{\max}=100$ to bound the operator state.

\noindent\textbf{Results.}
On both datasets (Figure~\ref{fig:micro-tradeoffs}), the LLM-based strategies place boundaries more
accurately than embeddings: M1 and M2 both have higher boundary F1 than
M3. M3, in turn, is the fastest, since it needs only one embedding per
tuple, and M1 is faster than M2, since each M2 call also generates the
updated summary. The Jev-based M4 is less accurate than the LLM-based
strategies but several times faster; on W2E it is on the Pareto frontier between M3 and M1, while on MiDe22 M3
dominates it. Because the best strategy depends on the stream and the
throughput target, the planner selects the implementation per workload
at runtime (Section~\ref{sec:dynamic-planning}).

\subsection{Semantic Group-By}

\noindent\textbf{Partitioning in Stream Processing.}
Where windowing splits a stream into consecutive finite scopes,
partitioning splits it by value. Relational \texttt{GROUP BY} collects tuples that agree on a
key; its streaming counterpart, \texttt{keyBy} in
Flink~\cite{carbone2015flink}, applies a key selector
$k = f(x)$ to each tuple and routes it to the partition
for its key, so downstream operators maintain per-key state.

To extend this to unstructured text, semantic operator systems such as
Lotus~\cite{patel2025semanticoperators} let an LLM decide which group
a record belongs to, replacing the extracted key with a judgment over
content. The decision is still made against a fixed group set,
computed once from the full corpus before grouping begins. Incremental and stream clustering
algorithms~\cite{Charikar2003IncrementalClustering,aggarwal2003clustream,cao2006denstream} evolve their
clusters online, but group by vector distance, so the
groups cannot follow a criterion stated in natural language.

In our setting the corpus is never complete, so there is no group set
to compute in advance. Consider a news feed covering an ongoing
conflict. When the query starts, the stories have not broken yet, so
there are no groups to name. Early ``peace talks'' and ``sanctions''
reports form separate groups, merge into a broader ``Ukraine
conflict'' group as the story unfolds, and are retired once coverage
moves on. The operator must discover the group set as the stream runs
and revise it as stories evolve, so $f$ reads not only the tuple but
also the groups that are live when it arrives.

\noindent\textbf{Definition.}
We propose \emph{dynamic semantic group-by}, which fills this gap by grouping
tuples semantically while letting the group set evolve. The operator
maintains a catalog of live groups $G_t = \{g_1, \dots, g_K\}$, each
summarized by a name and description, and routes each arriving tuple by
\[
\mathrm{assign}(x_t, G_{t-1}) \in G_{t-1} \cup \{g_{\text{new}}\},
\]
attaching the resulting group identifier to the tuple and emitting it
downstream. Unlike \texttt{keyBy}, the assignment reads the catalog as
well as the tuple, so it is stateful: the same tuple may be routed
differently depending on which groups are live when it arrives. Also
unlike \texttt{sem\_window}, the output is a stream rather than a
window specification, so a query may consume the
labeled stream directly or route it into a downstream operator that
maintains per-group state. Because the stream is
unbounded, we cap the catalog at $G_{\max}$ live groups: when a new
group would exceed the cap, the longest-idle group is retired, which bounds the operator's state.

\noindent\textbf{Implementation.}
We implement five approaches whose interfaces are shown in
Figure~\ref{fig:streaming-ops-api}.
\emph{M1 (LLM-based)} passes each tuple to the LLM together with the
current catalog, and the LLM either assigns it to an existing group or
creates a new one.
\emph{M2 (LLM with Refinement)} adds a background pass every $k$
tuples that merges, splits, or renames groups to track topic drift and writes the refined catalog back to the operator
state, so that subsequent assignments read the refined groups; tuples
already emitted keep their group identifiers. The pass
does not block the main pipeline.
\emph{M3 (Embedding-based)} uses leader
clustering~\cite{hartigan1975clustering}. Let $\mathbf{e}_t$ be the
embedding of $x_t$ and $\mathbf{c}_j$ the centroid of group $g_j$. The
tuple joins the most similar group if its cosine similarity is at least
$\tau$ and otherwise opens a new group, and the chosen centroid is
updated by its running mean,
$\mathbf{c}_{j^*} \leftarrow \mathbf{c}_{j^*} + (\mathbf{e}_t - \mathbf{c}_{j^*})/n_{j^*}$,
where $n_{j^*}$ is the group size after adding $x_t$. The LLM is called
only to name new groups. M3 is cheap and fully online, but its groups
depend on arrival order and are never merged.
\emph{M4 (Embedding with Refinement)} extends M3 by
re-clustering the group centroids every $k$ tuples, with spherical
$k$-means and the number of clusters chosen by silhouette score~\cite{rousseeuw1987silhouettes}, to merge groups that M3
splits apart.
\emph{M5 (Jev-based)} keeps M1's catalog-based assignment but replaces
the generative LLM with a Jev decision model. Each assignment is
posed as a multiple-choice question whose candidates are the live
groups plus a \emph{none of the above} option; Jev returns a
probability for each candidate, so the number of candidates it scores
grows with the catalog. If \emph{none of the above} is selected, a
generative LLM names the new group, as in M3.
Because concurrent requests in M1, M2, and M5 do not see each other's
catalog updates, they can over-split. We therefore merge textually similar groups by Jaccard
similarity.

\noindent\textbf{Evaluation.}
We compare the five approaches on MiDe22 and W2E, reporting throughput
and per-item F1, which scores each tuple's group assignment against its
event label (group-by column of Figure~\ref{fig:micro-tradeoffs}). The
adjusted Rand index (ARI)~\cite{hubert1985comparing} gives the same
ranking, so we omit it. We set $G_{\max}=100$; no group is retired
on either dataset. M2 and M4 refine every $k=10$ tuples, and all
LLM-based and Jev-based methods run at a request concurrency of 10, so
each batch of concurrent assignments reads the groups refined after the
previous batch.

\noindent\textbf{Results.}
LLM-based assignment is the most accurate, with M2 the best on both
streams and by far on W2E, while embedding-based assignment is an order
of magnitude faster, since M3 and M4 call the LLM only to name new
groups. Refinement improves both families (M2 over M1, M4 over M3), and
with it embeddings come close to unrefined LLM assignment: M4 matches
M1's F1 on MiDe22 and trails it by 0.07 on W2E, at about $20\times$
the throughput. Jev (M5) does not help here: although faster than the
LLM-based methods, it rarely selects \emph{none of the above}, so it
merges distinct events and is dominated by M4 on both datasets. As with
semantic windows, the best trade-off depends on the stream and
throughput target.

\begin{figure}[t]
 \centering
 \includegraphics[width=\linewidth]{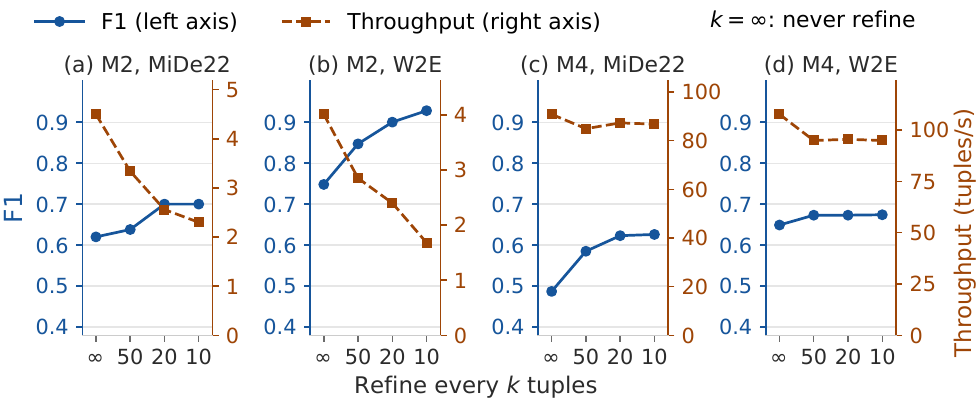}
 \caption{Refinement frequency for \texttt{sem\_groupby}, refining every $k$ tuples ($k=\infty$: never).
 Refinement generally improves F1, with smaller gains at higher frequency; its throughput cost
 is substantial for M2 and negligible for M4.}
 \label{fig:refine-freq}
\end{figure}

Figure~\ref{fig:refine-freq} further shows how the refinement interval
$k$ affects M2 and M4 at the same request concurrency of 10. More frequent refinement generally
improves F1, though the gain shrinks as refinement becomes more frequent.
Each M2 refinement is an LLM call over the catalog, so its
throughput drops as $k$ decreases, whereas M4 only re-clusters
group centroids without a model call, so its throughput
stays nearly flat.

\subsection{Continuous RAG}

\begin{figure}[t]
 \centering
 \includegraphics[width=\columnwidth, trim=10 10 10 80, clip]{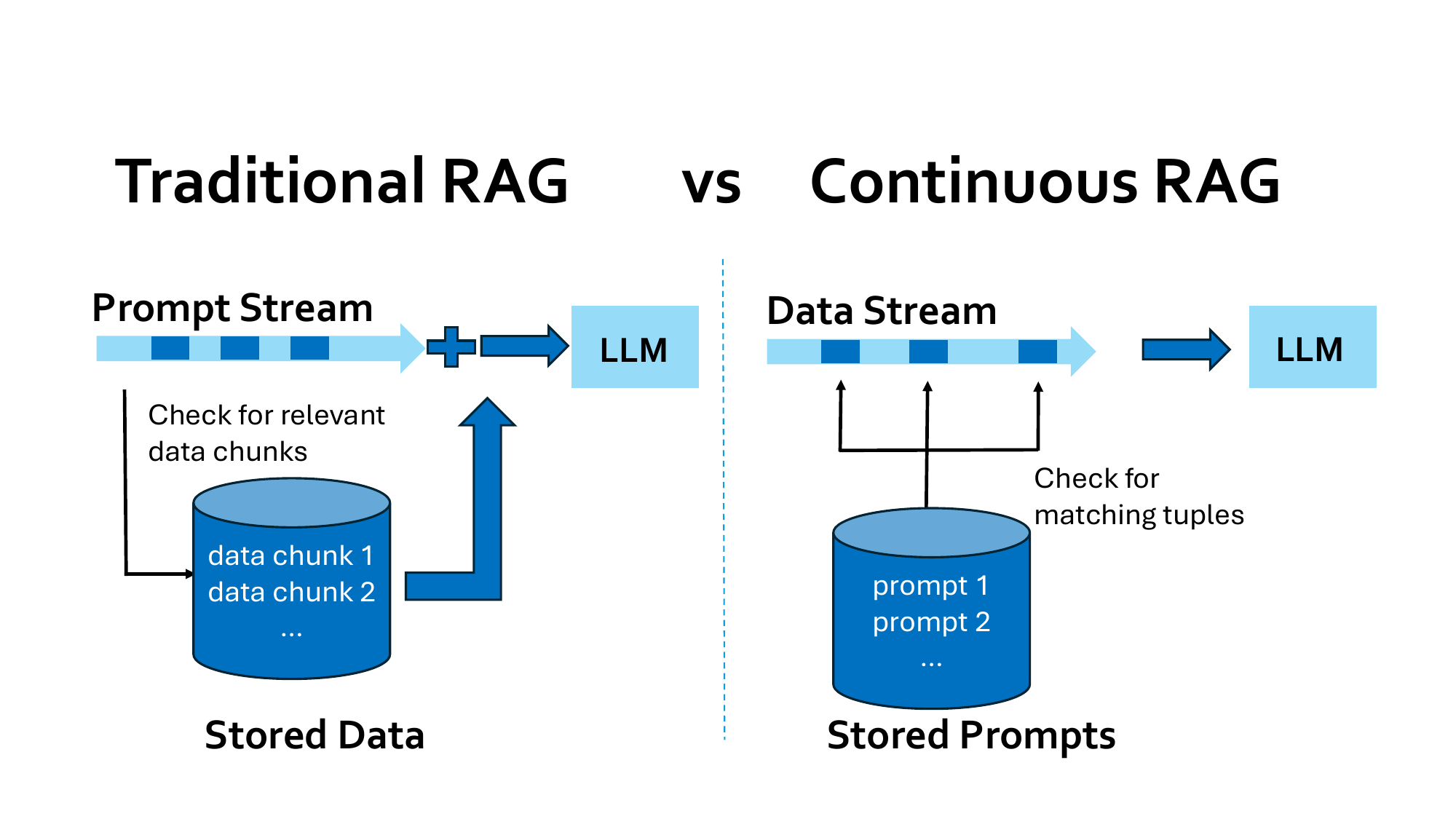}
 \caption{Traditional RAG (left) streams prompts against stored data;
 Continuous RAG (right) streams data against stored prompts,
 surfacing matches continuously without reissuing queries.}
 \label{fig:continuous-rag}
\end{figure}

To support retrieval over evolving streams, we introduce
\textit{continuous RAG}, the streaming analogue of traditional
Retrieval-Augmented Generation (RAG)~\cite{Lewis2020RAG}. As shown in
Figure~\ref{fig:continuous-rag}, the two paradigms are structural
inverses: in traditional RAG, prompts arrive as a stream and are
matched against a static data store; in continuous RAG, data arrives as
a stream and is matched against a persistent prompt store. This
inversion suits long-running monitoring tasks where retrieval intent is
stable but the data is unbounded. Continuous RAG has two parts: a
prompt store derived from a user's intent, and operators that match
each arriving tuple against that store. We build two operators on it,
\texttt{cts\_filter} and \texttt{cts\_topk}, and use one running example: a portfolio manager who holds TSLA, AAPL, and NVDA
and monitors a stream of financial news.

\noindent\textbf{Prompt Store.}
The user provides an \textit{intent} $I$, a natural-language request
that refers to a reference table $R$ of entities to monitor through a
placeholder such as \{stock\_portfolio\}.
Once the intent is registered, the LLM transforms it into the actual
matching prompt, a \textit{prompt template} $\hat{p}$: it reads the
intent and the schema of $R$, picks the relevant columns, and expands
the intent into a detailed instruction that takes those columns as parameters, so the user
never has to reference individual columns of $R$. Binding a row $r$ into the template yields its \textit{prompt
instance} $p_r = \hat{p}(r)$, so the LLM runs once per intent, not once
per entity. The prompt store $\mathcal{P}_t = \{\, p_r \mid r \in R_t
\,\}$ holds one instance per current row of $R$. It is a materialized
view over $R$~\cite{Gupta1995MV}, maintained incrementally without LLM
calls: buying or selling a holding adds or retires one instance.

\noindent\textbf{Continuous Filter (\texttt{cts\_filter}).}
\texttt{cts\_filter} emits every tuple that is relevant to at least one
entity:
\[
\mathrm{out} = \{\, x_t \mid \exists\, r \in R_t:\ \mathrm{match}(x_t, p_r) \,\}.
\]
In the running example, the manager registers \textit{``Find news that
would affect my \{stock\_portfolio\}''}, and the LLM produces a
matching template:
\begin{itemize}
 \item $\hat{p}$: \textit{``Does this article report news that could
 affect \{company\} (\{ticker\}), such as its earnings or guidance,
 its products, its executives, its suppliers or major customers, or
 regulation of its industry?''}
\end{itemize}
Unlike \texttt{sem\_groupby}, \texttt{cts\_filter} only filters: a downstream operator such as
\texttt{sem\_groupby} groups the kept tuples.

\noindent\textbf{Continuous Top-$k$ (\texttt{cts\_topk}).}
When matches arrive faster than a user or a downstream LLM can consume
them, \texttt{cts\_topk} keeps only the $k$ highest-scoring tuples per
entity within the enclosing window $W$:
\[
\mathrm{out}_r(W) = \operatorname*{top\text{-}k}_{x \in W} \; \mathrm{score}(x, p_r), \qquad r \in R_t.
\]
Because a stream has no end, top-$k$ must be scoped by a window. We use
a count-based window, though any window type applies, including semantic windows. Over the same portfolio,
the manager registers \textit{``Surface news most likely to move my
\{stock\_portfolio\}''} with $k=3$ and a window of 10 articles. The LLM then produces a scoring template:
\begin{itemize}
 \item $\hat{p}$: \textit{``Rate from 0 to 3 how likely this article
 moves the stock price of \{company\} (\{ticker\}); rate earnings, guidance, and regulatory actions high and routine commentary low.''}
\end{itemize}
 \texttt{cts\_topk} ranks articles and, at window close,
emits the top three for each holding. Ties are broken by string order.

\noindent\textbf{Implementation.}
We implement three single-stage matchers and three two-stage
cascades.
\textit{LLM} evaluates each (tuple, prompt instance) pair with one LLM
call, which returns a boolean for \texttt{cts\_filter} and a relevance
score for \texttt{cts\_topk}. Per-tuple cost therefore grows linearly
with $|R_t|$.
\textit{Jev} poses the same pairs to the Jev decision model, as a
yes/no question for \texttt{cts\_filter} and as a scored question over
the template's rating levels for \texttt{cts\_topk}. Like LLM, it
evaluates every pair, but without generating text.
\textit{Emb} embeds each prompt instance once on insertion, and matches tuples by cosine similarity. It emits a tuple
when the similarity reaches $\tau$ for \texttt{cts\_filter}, and uses
it as the score for \texttt{cts\_topk}. Each tuple needs one
embedding and $|R_t|$ cheap vector comparisons.

The cascades follow the classical filter-and-refine
strategy~\cite{Orenstein1986Spatial}: a cheap first stage keeps the $m$
prompt instances that best match each tuple, and a more accurate second
stage verifies or scores only these candidates. We combine the stages
as \textit{Emb$\rightarrow$LLM}, \textit{Emb$\rightarrow$Jev}, and
\textit{Jev$\rightarrow$LLM}. The second stage runs at most $m$ times
per tuple, independent of $|R_t|$. Recall, however, is bounded by the
first stage, because pruned pairs never reach the second; in
particular, a tuple relevant to more than $m$ entities loses its excess
matches.

\noindent\textbf{Evaluation.}
We evaluate both operators on MiDe22 and W2E, using a small subset of
each dataset's events as the reference table. F1 for
\texttt{cts\_filter} uses the datasets' event labels, and Recall@10
for \texttt{cts\_topk} uses the gold model's top-10. For \texttt{cts\_topk} we set $k=10$
with a count-based window of 50 tuples, and all cascades forward the
top $m=3$ candidates. Results are shown in Figure~\ref{fig:micro-tradeoffs}.

\noindent\textbf{Results.}
On these datasets, Jev is faster than the LLM and more accurate on
the yes/no decisions of \texttt{cts\_filter}, but loses recall on the
graded ranking of \texttt{cts\_topk}. Jev does not pay off as a
first stage: Jev$\rightarrow$LLM is dominated on every dataset. An
embedding first stage does: it makes Emb$\rightarrow$LLM and
Emb$\rightarrow$Jev \mbox{1.5--3.4$\times$} faster than their second
stage alone, at a loss of at most 0.03 in F1 or Recall@10. Embeddings alone are
the fastest but least accurate matcher: similarity is good enough to
shortlist entities, but not to decide. Since the best matcher differs
between the two operators, the planner (Section~\ref{sec:dynamic-planning})
selects it per operator.
\section{Dynamic Planning Framework}
\label{sec:dynamic-planning}

Section~\ref{sec:streaming-ops} showed that every streaming operator
offers several implementations with different throughput--accuracy
trade-offs, and that no implementation is best on every stream. The right choice also depends on load. When the arrival
rate rises, an accurate but slow plan falls behind the stream; when it
drops, a fast plan leaves accuracy on the table. A CP therefore cannot
fix its plan in advance; it must choose the plan that fits the current stream and the current load.

\begin{figure*}[t]
 \centering
 \includegraphics[width=\textwidth,height=0.3\textheight,keepaspectratio]{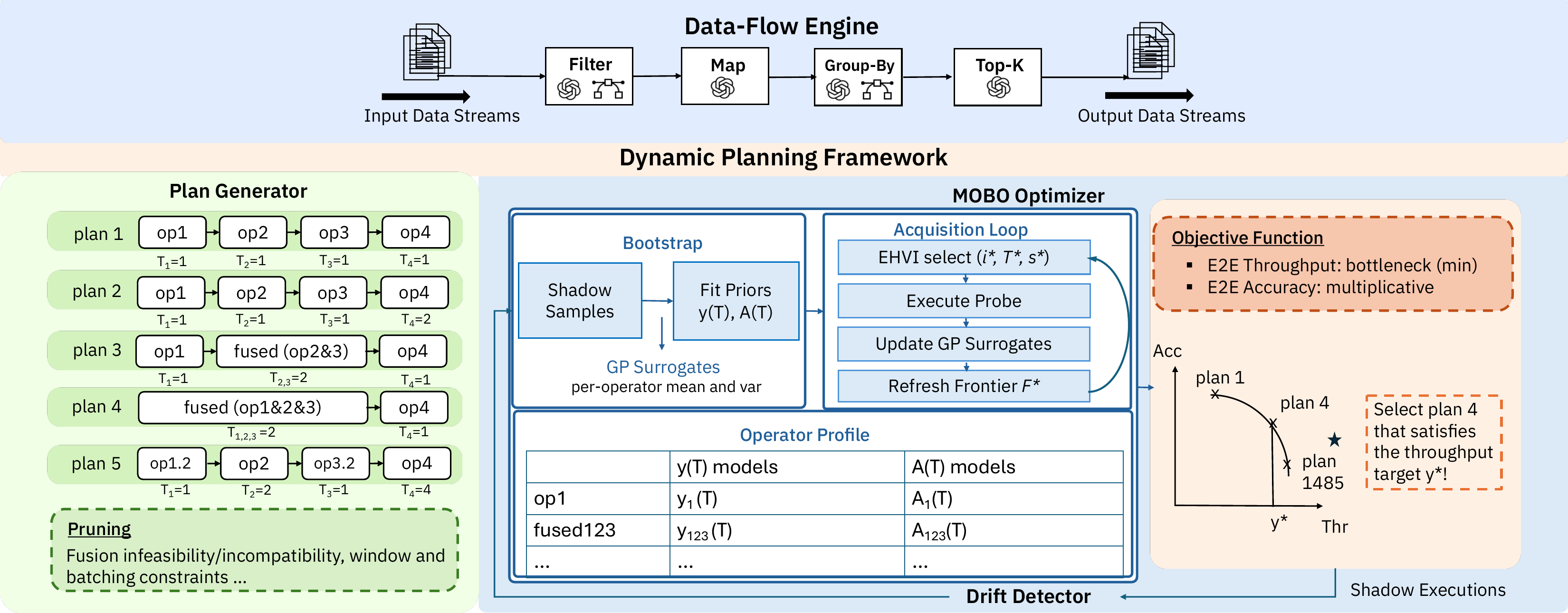}
 \caption{Overview of the dynamic planning framework: (1) a plan
 generator that enumerates and prunes candidate plans, (2) a
 cost-aware optimizer that learns the Pareto frontier under a limited
 probing budget, and (3) an online adapter that tracks load along the
 frontier and triggers re-optimization under drift.}
 \label{fig:dynamic-planning-framework}
\end{figure*}

We formalize this as follows. Given a logical query plan over a stream with
arrival rate $\lambda(t)$, the planner chooses at each point in time a
physical plan $\pi$ from a space $\Pi$ of candidate configurations
that maximizes end-to-end accuracy $A_{\mathrm{e2e}}(\pi)$ subject to
 $y_{\mathrm{e2e}}(\pi) \ge \lambda(t)$. Monitoring applications must
process every tuple, so the planner never drops tuples to keep up. A plan that falls behind
accumulates an unbounded backlog no matter how accurate it is, so the set
of feasible plans shrinks as load rises. Because $\lambda(t)$ is not
known in advance, the planner does not solve this problem for a single
target rate. Instead, it learns the full throughput--accuracy Pareto
frontier of the plan space, and at runtime selects the most accurate
point on that frontier that sustains the current rate.
Figure~\ref{fig:dynamic-planning-framework} shows the resulting
framework. A \emph{plan generator} enumerates and prunes candidate
configurations (Section~\ref{sec:plan-space}); a \emph{cost-aware optimizer}
learns the frontier under a limited probing budget using per-operator
performance models (Section~\ref{subsec:cost-aware-estimation}); and an
\emph{online adapter} follows changes in load along the frontier and
refreshes it when the stream drifts (Section~\ref{sec:online_mobo}).

\subsection{Plan Space}
\label{sec:plan-space}

 Of the many known optimizations for semantic
operators~\cite{patel2025semanticoperators,russo2025abacus,wei2025moar},
our planner focuses on choices
that trade accuracy for throughput most directly on fixed hardware, and
searches over four dimensions jointly: (1)~the \emph{operator
implementations} of Section~\ref{sec:streaming-ops};
(2)~\emph{model selection}, which assigns each generative operator a
model from a candidate pool; (3)~\emph{tuple batching}, which packs $T$ tuples
into a single LLM call to amortize the instruction and per-call
overhead, and is known in the LLM literature as batch
prompting~\cite{cheng2023batch,Lin2024BatchPrompt}; and (4)~\emph{operator
fusion}, which executes adjacent operators in a single call so that
their shared input is encoded once.

The last two dimensions are the LLM counterparts of classical
amortization, but, unlike in a relational engine, they are not
semantics-preserving. With tuple batching, each tuple is judged in the
presence of the others, and LLM accuracy is known to be sensitive to the
composition and order of items in a
prompt~\cite{lu2022fantastically,zhao2021calibrate,liu2024lost}. With
fusion, instructions for different operators share one prompt, and a
fused call generates output for every tuple even when a downstream
filter would have discarded it. Both raise throughput at a
possible cost in accuracy, which makes them planning decisions rather
than free rewrites. Batching also delays a tuple until its batch fills,
by at most $(T-1)/\lambda$ at arrival rate $\lambda$. The planner
optimizes sustained throughput; applications with a latency bound can
impose it by capping $T$. Since the accuracy cost of both
transformations varies by operator and data, the planner makes no
assumption about it: it measures accuracy at sampled batch sizes, and
measures each fused chain. The plan generator enumerates the cross product
of these dimensions and prunes configurations that are infeasible, such
as fusions of incompatible operators or batch sizes that violate window
constraints.

\noindent\textbf{Threshold Calibration.}\phantomsection\label{par:threshold-calibration}
We calibrate each implementation's decision
threshold $\tau$ offline on a held-out stream sample. We sweep $\tau$
over $[0, 1]$ in steps of 0.05, keep the value that maximizes accuracy,
and fix it during MOBO exploration; its cost counts toward the MOBO probing budget.

\subsection{Frontier Learning}
\label{subsec:cost-aware-estimation}

Exhaustively evaluating every plan configuration is infeasible. The central
challenge is therefore deciding \emph{which} operators to probe, at
\emph{which} batch sizes, and at \emph{what} sampling rates, so as to
improve the frontier estimate as much as possible under a fixed budget
$B$. We formalize this as budget-constrained multi-objective Bayesian
optimization (MOBO)~\cite{shahriari2016bayesopt,daulton2020differentiable}: using
$N$ probes, where the $n$-th probe $(i_n, T_n, s_n)$ runs operator
$i_n$ at batch size $T_n$ and sampling rate $s_n$, identify the Pareto
frontier
\begin{equation}
 \begin{aligned}
&\mathcal{F}^\ast = \operatorname*{Pareto}_{\pi \in \Pi}
\bigl(y_{\mathrm{e2e}}(\pi),\, A_{\mathrm{e2e}}(\pi)\bigr) \\
&\text{s.t.}\quad
\textstyle\sum_{n=1}^{N} \mathrm{cost}(i_n, T_n, s_n) \le B,
\end{aligned}
\label{eq:mobo}
\end{equation}
where $\Pi$ is the space of feasible plan configurations,
 $y_{\mathrm{e2e}}(\pi)$ and $A_{\mathrm{e2e}}(\pi)$ denote end-to-end
throughput and accuracy, and each probe consumes part of the
budget $B$. Probes act on individual operators rather than whole plans:
plan-level predictions are composed from operator-level models, so a
single probe refines every plan that contains the probed operator.

Abacus~\cite{russo2025abacus} also probes individual physical
operators, but treats each as an independent arm of a multi-armed
bandit and optimizes each logical operator separately. Our planner
differs in two ways. First, configurations of one variant are related:
applied here, each batch size would be a separate arm, yet throughput
at every batch size is governed by the same two serving costs. Our
surrogates model each variant as a function of $T$, so a few probes
predict throughput at every batch size, and accuracy observations carry
over to nearby batch sizes. Second, we choose each probe by its
expected gain in end-to-end throughput and accuracy, rather than by how
much it improves the estimate of a single operator, as we describe
next.

\noindent\textbf{Operator Models.}
Each operator $i$ is modeled by a pair of Gaussian process (GP)
surrogates~\cite{rasmussen2006gaussian}, probabilistic models that
approximate an expensive function from a small number of observations,
one for throughput and one for accuracy, with predictive means
$\mu_{y_i}(T)$ and $\mu_{A_i}(T)$ and variances $\sigma^2_{y_i}(T)$ and
$\sigma^2_{A_i}(T)$ as functions of batch size $T$. Fused variants
(e.g., $y_{123}$ and $A_{123}$ for the fusion of operators 1--3),
implementation alternatives (e.g., $y_{1.2}$ and $A_{1.2}$ for the second
implementation of operator~1), and model alternatives are treated as
standalone operators with their own surrogates, since their throughput
and accuracy cannot be derived from those of other variants.

Throughput follows from how the serving engine processes requests,
which gives its GP a structural mean. $P$ concurrent calls of $T$
tuples each take
\begin{equation}
L(T,P) \;=\;
P\,(n_0 + n_1 T)\,t_{\mathrm{in}}
\;+\; (m_0 + m_1 T)\,(t_{\mathrm{w}} + P\,t_{\mathrm{kv}})
\;+\; d,
\label{eq:latency}
\end{equation}
where a call has $n_0$ input and $m_0$ output tokens of its own, each
tuple adds $n_1$ and $m_1$, $t_{\mathrm{in}}$ is the prefill time per
token, each decode step reads the weights once ($t_{\mathrm{w}}$) and
each call's KV cache ($t_{\mathrm{kv}}$), and $d$ is a fixed overhead.
With $P$ fixed by the deployment, the time per tuple reduces to
\begin{equation}
1/y(T) = L(T,P)/(PT) = a + b/T,
\label{eq:throughput}
\end{equation}
where $a$ collects per-tuple and $b$ per-call costs. We fit $a$ and $b$
from a few probes and use the curve as the GP mean; the GP residual
absorbs what the model omits.

Accuracy has no comparable structure, since the effect of batching and
fusion varies by operator and data. Its GP
therefore uses the standard constant mean~\cite{balandat2020botorch} and
learns how accuracy varies with $T$ entirely from measurements, taken as
agreement with the gold model, GPT-4o, as in
 Section~\ref{sec:streaming-ops}. Finally,
probes on smaller samples give noisier estimates, so each observation is
assigned a noise variance proportional to $1/s$, where $s$ is the
sampling rate.

The two objectives are composed from operator-level predictions through
composition functions, and we treat operators as conditionally
independent given their surrogates. Operators execute in a pipeline, so
the slowest operator is the bottleneck and determines end-to-end
throughput:
\begin{equation}
 y_{\mathrm{e2e}}(\pi) = \min_i \frac{y_i(T_i)}{\phi_{<i}},
\qquad \phi_{<i} = \prod_{j<i} \rho_j,
\label{eq:e2e-throughput}
\end{equation}
where $\phi_{<i}$ is the fraction of the stream that reaches operator
$i$, given the selectivities $\rho_j$ of the operators before it. An
operator that sees only part of the stream can therefore be slower
without becoming the bottleneck. Selectivities are measured per variant
during probing and refreshed at runtime from live operator counters.
This composition assumes that operators on different model servers do
not contend (Section~\ref{sec:vectraflow}); for models that share a GPU,
it is an approximation.
End-to-end accuracy is the agreement between a plan's final output
and that of a gold pipeline that runs every operator with
GPT-4o~\cite{openai2024gpt4o} (Section~\ref{sec:experimental-setup}).
It is approximated as
the product of per-operator accuracies,
 $A_{\mathrm{e2e}}(\pi) \approx \prod_i A_i(T_i)$, as in prior LLM pipeline systems~\cite{russo2025abacus}; this approximation assumes that
operators err independently. Fused chains are measured
directly, so independence is assumed only across separately executed
operators. Users may supply custom
composition functions for other pipeline topologies and for pipelines
whose accuracy dependencies are known, and the planner falls back to sparse
pipeline-level sampling when shadow executions reveal a significant
deviation between the composed prediction and the observed end-to-end
accuracy.

 \noindent\textbf{Bootstrap Profiling.}
Before probing, the optimizer needs a coarse estimate of every
operator's throughput and accuracy curves. Each operator is therefore
probed at a small sampling rate (e.g., $s_0{=}0.1$) and a few
representative batch sizes (e.g., $T \in \{1,2,8\}$). These
measurements fit the throughput mean function of
Eq.~\eqref{eq:throughput} and give each accuracy GP its first
observations, so that later probes refine the surrogates rather than
rediscover their coarse shape.

\noindent\textbf{Search.}
Each iteration chooses the next probe by maximizing a cost-aware
acquisition function that balances expected frontier improvement
against probing cost, extending expected improvement per
second~\cite{snoek2012practical} to the multi-objective setting:
\begin{equation}
(i^\ast, T^\ast, s^\ast) = \arg\max_{i,T,s}\;
\frac{\mathrm{EHVI}(i,T,s)}{\mathrm{cost}(i,T,s)}.
\label{eq:acquisition}
\end{equation}
The numerator measures how much a probe is expected to improve the
frontier. Expected Hypervolume Improvement
(EHVI)~\cite{daulton2020differentiable,daulton2021parallel} quantifies
the expected increase in the area dominated by the
frontier $\mathcal{F}$, relative to a reference point $r$:
\begin{equation}
\mathrm{EHVI}(i,T,s) = \mathbb{E}\!\left[
\mathrm{HVI}\bigl((y_{\mathrm{e2e}}, A_{\mathrm{e2e}}),\,\mathcal{F};\,r\bigr)
\right].
\end{equation}
Because the composed end-to-end predictions are not Gaussian, we
evaluate the expectation by Monte Carlo, drawing samples from the
operator posteriors and composing them per plan. The denominator
penalizes expensive probes by estimating their execution time,
\begin{equation}
\mathrm{cost}(i,T,s) = \frac{n s}{\hat{y}_i(T)},
\end{equation}
where $n$ is the number of profiling tuples, $s$ the sampling rate, and
$\hat{y}_i(T)$ the posterior mean throughput, so that cheaper and faster
probes are preferred when they offer comparable improvement. After
executing probe $(i^\ast, T^\ast, s^\ast)$, we update the operator's GP
surrogates, refresh the frontier $\mathcal{F}$, and repeat until the
budget $B$ is exhausted. The final Pareto frontier is computed from surrogate posterior means across all explored configurations and used for runtime adaptation.

\subsection{Online Adaptation}
\label{sec:online_mobo}

The offline MOBO phase produces a frontier calibrated to the
distribution seen during profiling. At runtime, the adapter deploys the
most accurate frontier plan whose throughput $y_{\mathrm{e2e}}(\pi)$ is
at least the arrival rate $\lambda$, measured over the last $W$ tuples,
and reselects when $\lambda$ changes, without probing. When the stream
shifts, it detects which operators have drifted, re-optimizes them,
and reselects from the refreshed frontier.

\noindent\textbf{Shadow Execution.}
The surrogates predict each operator's throughput and accuracy; shadow
execution measures them for the deployed plan. Throughput is read from the plan's execution at no extra cost, as
tuples processed per second of busy time, during which the operator has
calls in flight, so that it measures capacity rather than the arrival rate. Live tuples carry no labels, so
for accuracy, every $W$ tuples the system samples a fraction $s$ of
them and runs the gold model, GPT-4o, on each operator's input.
The gold model runs on separate resources of the user's choosing, such as
a hosted API or a locally deployed model, so it does not consume the
pipeline's GPUs or count against its throughput.
Comparing per operator rather than end-to-end localizes drift to
individual operators.

\noindent\textbf{Surrogate Drift Detection.}
Each operator $i$ carries GP surrogates $\mathcal{GP}_i^y$ and
$\mathcal{GP}_i^A$ from the offline phase. Given a shadow observation
$(y_i^{\mathrm{obs}}, A_i^{\mathrm{obs}})$ at batch size $T$, we compute
standardized residuals
\begin{align}
 r_i^y &= \frac{y_i^{\mathrm{obs}} - \mu_i^y(T)}{\sigma_i^y(T)}, \quad
 r_i^A = \frac{A_i^{\mathrm{obs}} - \mu_i^A(T)}{\sigma_i^A(T)},
\end{align}
where each predictive variance includes observation noise. We
feed each residual sequence to a Page-Hinkley
test~\cite{gama2014survey}, which accumulates deviations across windows
and flags operator $i$ when the accumulated deviation in either
 objective exceeds a threshold, calibrated on a drift-free stream. Small
but persistent shifts trigger re-optimization; isolated noisy
windows do not.

\noindent\textbf{Re-Optimization and Plan Update.}
When operator $i$ is flagged, the planner runs a targeted MOBO pass on
it under a budget $B_{\mathrm{online}} \le B$, warm-started from the
offline surrogates. Since a content shift affects the logical operator
rather than one implementation, the pass covers all of operator $i$'s
variants: it adds the shadow observations to their surrogates and
inflates the variance of the offline observations so that fresh probes
dominate, while all other operators keep their surrogates. We do not
rebuild from scratch, since drift moves an operator's curves more than
it reshapes them, and the offline data still carry their shape. The
planner then resets the operator's Page-Hinkley statistics, recomputes
the Pareto frontier, and selects the most accurate plan on it that
sustains the current rate.

\noindent\textbf{Plan Switching.}
Since CP operators are stateful, switching a plan can also change the
state an operator keeps: switching a semantic window from a running
summary to an embedding centroid requires state conversion, replay, or
waiting for a window boundary. Prior work on stream engines migrates plans by
converting the old state into the new plan's
state~\cite{zhu2004migration,yang2007hybmig} or by running both plans
until the old state expires~\cite{zhu2004migration,kraemer2006migration}.
We avoid conversion and replay by switching only where the old state
is empty or portable. For the window-scoped operators
(\texttt{sem\_window}, \texttt{sem\_agg}, \texttt{sem\_topk},
\texttt{cts\_topk}), the new plan starts when the current window
closes. For the operators with state across windows
(\texttt{sem\_groupby}, \texttt{cts\_filter}), the new plan reuses the
existing state, the group catalog or the prompt store; an embedding-based implementation simply embeds it first,
as it does when a group or prompt is added. Batch size, fusion, and the model behind an LLM-based
implementation leave the state untouched and switch at the next call.

\section{Evaluation}
\label{sec:evaluation}

In addition to the operator-level microbenchmarks presented earlier,
we evaluate CPs end-to-end on two representative streaming
pipelines to answer four system-level questions:

\begin{itemize}[leftmargin=1em]
    \item \textbf{How much do execution optimizations improve end-to-end performance?}
We evaluate how much the set of execution optimizations improves pipeline throughput, and how much accuracy each speedup costs.

    \item \textbf{How efficiently can MOBO recover Pareto frontiers?}
    We measure how effectively cost-aware sampling guided by MOBO
    identifies Pareto-efficient plans compared with heuristic
    exploration baselines.

    \item \textbf{How well do CPs adapt online under workload and distribution shifts?}
    We examine whether CPs can dynamically traverse the
    precomputed Pareto frontier under increasing arrival rates, and
    whether their drift detector with targeted re-optimization can
    efficiently recover stale frontier regions under changing data
    distributions.

    \item \textbf{How effective is each optimization?}
    We ablate Jev-based implementations,
    tuple batching and fusion, and model selection, and compare each frontier with the full set of optimizations and with the
    unoptimized baseline.
\end{itemize}

\subsection{Experimental Setup}
\label{sec:experimental-setup}

\noindent\textbf{Hardware and Model.}
We evaluate CPs using Qwen2.5-7B and Qwen2.5-14B~\cite{yang2025qwen}, Llama-3.1-8B~\cite{grattafiori2024llama}, and Open-Jev-2B~\cite{openjev}, deployed via the vLLM~\cite{kwon2023pagedattention} inference server. All end-to-end experiments were run on three NVIDIA L40S GPUs (48\,GB each). Temperature is 0 unless otherwise noted. One GPU hosts Qwen2.5-7B and Open-Jev-2B, one Qwen2.5-14B, and one Llama-3.1-8B. Each server uses prefix caching and concurrent requests to saturate its GPU.

\noindent\textbf{Pipelines and Datasets.}
\emph{Stock News Monitoring.}
We replay financial news from the FNSPID~\cite{dong2024fnspid} dataset as a stream. A \texttt{cts\_filter} keeps articles relevant to a given stock
portfolio. The filtered stream passes to \texttt{sem\_map},
which performs data cleaning and structuring. Next,
\texttt{sem\_groupby}
in static mode routes articles to the portfolio's company tickers
(e.g., TSLA, AAPL), and within each group \texttt{sem\_topk}
selects the most impactful articles within a count-based window of size 10, producing a feed of the most
impactful news per company.

\noindent\emph{Misinformation Event Monitoring.}
We continuously ingest the MiDe22 stream~\cite{toraman2024mide22}, which is temporally ordered
across 40 distinct real-world events spanning topics with abrupt transitions between events and gradual semantic evolution within each, making it a natural testbed for content drift. The stream mixes misinformation with accurate tweets, and a \texttt{sem\_filter} discards the misinformation. The remaining
tuples are grouped by semantic topic via
 \texttt{sem\_groupby} in dynamic mode, which tracks shifting topic distributions
as the stream transitions across events. Each topic flow is segmented by \texttt{sem\_window}, which opens a new window at detected event boundaries so the downstream \texttt{sem\_topk} always selects the three most urgent items within a semantically coherent event scope.

\noindent\textbf{Metrics.}
We report sustained throughput (tuples/s), end-to-end accuracy (F1), and
p95 latency.
End-to-end accuracy is the F1 score of each pipeline's final output
against a gold pipeline that runs the same query with GPT-4o at every
operator.
It does not compare window boundaries or group identifiers
directly, since they are intermediate results that change dynamically.
Windowing and grouping errors show up in the output: they change
which tuples the final top-$k$ selects, so the output accumulates the
errors of every upstream operator. Splitting a window adds outputs, and merging two drops outputs; both lower F1.
Operator-level accuracy, which the microbenchmarks report
(Section~\ref{sec:streaming-ops}) and the planner composes into its
estimate (Section~\ref{subsec:cost-aware-estimation}), is
measured against the dataset's labels or the gold model
on the same input.

\noindent\textbf{Comparison with Existing Systems.}
Batch semantic systems (e.g., Lotus~\cite{patel2025semanticoperators},
Palimpzest~\cite{liu2025palimpzest}) have no counterpart for
\texttt{sem\_window}, dynamic \texttt{sem\_groupby}, or the continuous
RAG operators, and they choose one plan per dataset rather than
re-planning as load and content change, so they cannot run our pipelines
as written. Palimpzest's optimizer, Abacus~\cite{russo2025abacus},
searches a different space (e.g., mixture of agents, prompt
optimization) from ours. We therefore compare against a production stream processor
and alternative planning strategies.

\subsection{Optimization Effectiveness}
\label{sec:optimization-effectiveness}

\noindent\textbf{Setup.} To construct the ground-truth
Pareto frontier, we evaluate more than $100{,}000$ configurations end-to-end on a sample of each stream, pruning unpromising plans after an
initial round; the resulting frontier is therefore an estimate. We report all improvements relative to the baseline below.

\noindent\textbf{Baseline.}
Our baseline, built on Qwen2.5-7B-Instruct~\cite{yang2025qwen}, reproduces how a production stream processor runs LLM
pipelines today: as in Apache Flink's \texttt{ML\_PREDICT}~\cite{flink22},
every operator uses its LLM-based implementation and issues one call per
tuple, with no tuple batching, fusion, implementation or model selection,
or accuracy-aware planning. It runs on the same three GPUs as CPs,
with one vLLM replica of Qwen2.5-7B per GPU, requests spread across the
replicas, and each GPU saturated. To confirm that this baseline
is representative of a production engine rather than an artifact of our
prototype, we run both pipelines on Flink~2.2 against the same
vLLM servers with identical requests. On both
pipelines, Flink and VectraFlow achieve roughly the same throughput and
accuracy (about 18.2~tuples/s at 0.771 on Stock News and 17.0~tuples/s
at 0.73 on Misinformation), so the
workload is bound by LLM inference rather than engine overhead.

\begin{figure}[t]
  \centering
  \includegraphics[width=\linewidth]{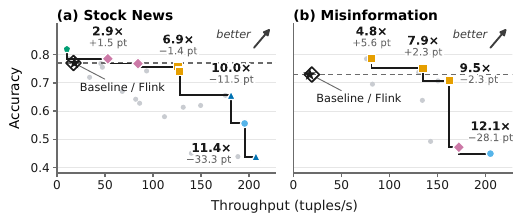}
  \caption{Pareto-optimal plans of both pipelines. Each
    marker shape and color is one plan structure (choice of operator
    implementations). The black staircase is the frontier, gray points are
    dominated plans measured in the ablation (Figure~\ref{fig:pareto-ablation}),
    the star and dashed line mark the unoptimized baseline, the hollow
    diamond marks Flink, and labels give the speedup and accuracy change
    (in points) over the baseline.}
  \label{fig:stepwise_adoption_combined}
\end{figure}

\noindent\textbf{Results.}
Figure~\ref{fig:stepwise_adoption_combined} shows the Pareto-optimal
plans of both pipelines.
\emph{(1) Moderate speedups cost no accuracy; larger ones do.} On both
pipelines, a plan beats the baseline on both axes (0.786 at $2.9\times$
and $4.8\times$). Within 2\% and 4\% of the baseline's accuracy,
throughput rises to $6.9\times$ and $9.5\times$, while the fastest
plans reach $11.4\times$ and $12.1\times$ but drop to 0.44 and 0.45
accuracy.
\emph{(2) Every plan applies batching or fusion,} but only with small
batches and fusions of simple operators (\texttt{sem\_map} and
\texttt{sem\_groupby}), which cost little or no accuracy;
larger batches and complex fusions lose accuracy and are dominated.
\emph{(3) Most plans also change implementations:} 88\% of Stock News plans and
all Misinformation plans change at least one operator's implementation.

\subsection{Frontier Learning Quality and Efficiency}
\label{sec:frontier-learning}

\begin{figure}[t]
  \centering
  \includegraphics[width=\linewidth]{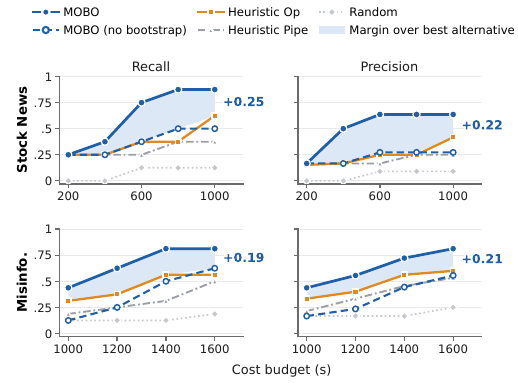}
  \caption{Recall and precision under increasing cost budget. The
  shaded band, labeled at the largest budget, is MOBO's margin over the best alternative.}
  \label{fig:budget}
\end{figure}

\noindent\textbf{Setup.}
We use 1K points to train throughput and accuracy models and 1K points to evaluate how each method recovers the ground-truth Pareto frontier from Section~\ref{sec:optimization-effectiveness}. The training set also serves as the calibration set for pruning fused candidates and for hyperparameter tuning, and is disjoint from the 1K evaluation set. Budgets are in seconds of probing. We compare five strategies: (1)~\emph{Random}, which profiles randomly sampled pipeline configurations; (2)~\emph{Heuristic Pipe}, which profiles a sample of full pipeline configurations and fixes the most promising strategy on that sample; (3)~\emph{Heuristic Op}, which does the same for each operator independently; (4)~\emph{MOBO (no bootstrap)}, our framework without bootstrap profiling; and (5)~\emph{MOBO}, the full framework.

\noindent\textbf{Results.}
Figure~\ref{fig:budget} shows how much of the ground-truth frontier
each strategy recovers as the probing budget grows.
\emph{(1) MOBO recovers the frontier with the least budget.} It matches
or beats every other strategy in recall and precision at every budget
on both pipelines, and recovers 88\% of the Stock News Pareto plans by
800~s.
\emph{(2) Probing operators beats probing pipelines.} One operator probe
improves the prediction for every plan that contains the operator, which
also lets Heuristic Op beat Heuristic Pipe.
\emph{(3) Cost-aware search and bootstrap profiling avoid wasted probes.}
MOBO picks each probe by expected frontier gain per unit cost, so it
spends little on operators that cannot reach the frontier, and with a
coarse model of every operator from bootstrap profiling, later probes
refine it instead of rediscovering it.
\emph{(4) Composition ranks plans correctly.} Although MOBO builds its frontier from composed operator-level predictions, it matches the measured frontier.

\subsection{Online Adaptation}
\label{sec:online-adaptation}

The offline Pareto frontier learned in
Section~\ref{sec:frontier-learning} serves two online roles.
First, it enables adaptation to arrival-rate
changes without
re-learning: at runtime, the system selects the most accurate frontier
plan that sustains the observed $\lambda$. Second, when the frontier becomes stale due
to distribution shift, CPs trigger targeted re-probing
to correct only the operators that have drifted. We evaluate
both scenarios in turn.

\begin{figure}[t]
  \centering
  \includegraphics[width=\linewidth]{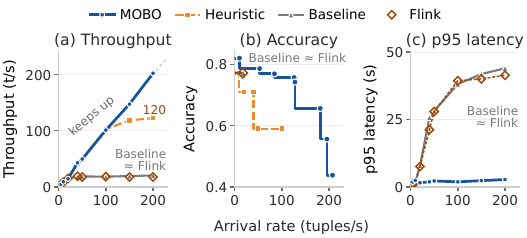}
  \caption{ Plan selection under increasing arrival rate on the Stock
    News pipeline. (a) Sustained throughput; the diagonal marks keeping
    up with arrivals. (b) Accuracy, plotted only up to the rate each method sustains. (c) p95 latency of MOBO, the baseline, and Flink.}
  \label{fig:real-streaming}
\end{figure}

\begin{figure}[t]
  \centering
  \includegraphics[width=\linewidth]{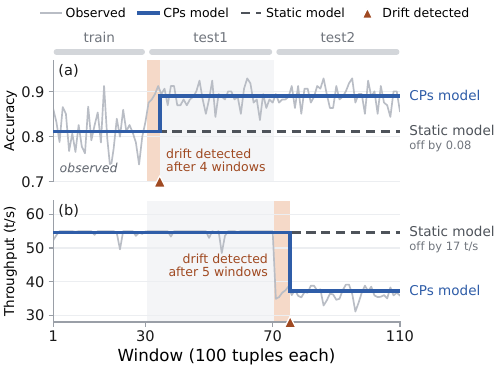}
  \caption{Accuracy and throughput of the Stock News filter
    under distribution shift: observed (gray), and as
    predicted by the CPs and static models. The shaded band runs from each shift to its detection (triangle).}
  \label{fig:drift-timeline}
\end{figure}

\noindent\textbf{Adaptation under Increasing Load.}
We replay the Stock News pipeline as a stream with Poisson inter-arrival
 times at fixed arrival rates $\lambda$ from 5 to 200~tuples/s, averaging
three runs per rate. In each run, MOBO deploys the plan it selects for that rate
from the Pareto frontier learned in Section~\ref{sec:frontier-learning}. These synthetic rates stress-test the planner. We compare MOBO with a \emph{static baseline} (fixed plan at deployment) and a \emph{heuristic} that selects cheaper configurations as load increases.

For each arrival rate, Figure~\ref{fig:real-streaming} reports the
sustained throughput (a), the accuracy (b), and
the p95 latency (c).
\emph{(1) Fixed plans fall behind early.} The static baseline and Flink
sustain fewer than 20~tuples/s; beyond that, their queues grow and their
p95 latency climbs to tens of seconds.
\emph{(2) MOBO keeps up across the entire range} (a), giving up accuracy
only when the arrival rate demands a faster plan and stepping down the
frontier one plan at a time.
\emph{(3) MOBO beats the heuristic on both axes.} The heuristic
saturates near 120~tuples/s and picks cheap plans too early: at every
rate where it keeps up, MOBO is more accurate.
\emph{(4) Batching adds little latency} (c). A tuple waits at most
$(T{-}1)/\lambda$ for its batch to fill; with at most 10 tuples per
batch, this is under 1~s once $\lambda \ge 10$~tuples/s, so MOBO's p95 latency stays flat at a few seconds.

\noindent\textbf{Adaptation under Distribution Shift.}
We train the planner on Stock News filtered for TSLA and AAPL, then, in
a single run, switch the live stream to two held-out distributions.
Test1 contains only TSLA news, which shifts the filter's accuracy but
not its input size. Test2 keeps TSLA news but nearly doubles the text
length, which shifts throughput. We stream 30 windows of training data
and 40 windows of each test at $\lambda{=}55$~tuples/s.
CPs run the adaptation loop every window of $W{=}100$ tuples with
$s{=}0.2$, use a Page-Hinkley test with $\delta{=}0.5$ and threshold 5,
and re-optimize under an online budget;
\emph{Static} keeps the offline surrogates and plan.

Figure~\ref{fig:drift-timeline} shows the filter's observed accuracy
and throughput with the values predicted by the CPs and static models.
\emph{(1) Detection is fast and targeted.} The detector catches each
shift within five windows and flags only the objective that moved:
accuracy at test1, where input size is unchanged, and throughput at
test2, where longer inputs slow the filter but leave its accuracy
unchanged.
\emph{(2) Re-optimization keeps estimates current.} After each targeted
re-optimization, the CPs model matches the observed filter again.
Static keeps its offline estimates and, at test2, still predicts that
the filter keeps up with $\lambda$, so it never notices that it has
fallen behind.
\emph{(3) Recovery is cheap, since it re-probes only the drifted operator.} Re-optimization recovers 67\% of the
Pareto plans on test1 and 80\% on test2 at $B_{\mathrm{online}}{=}100$~s,
and 83\% and 100\% at 200~s.

\subsection{Ablation Studies}
\label{sec:model-selection}

\begin{figure}[t]
    \centering
    \includegraphics[width=\columnwidth]{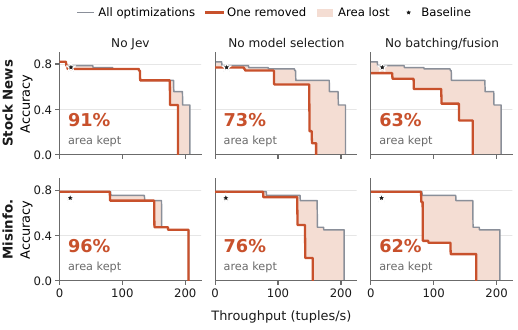}
    \caption{Ablation on both pipelines. Each panel removes one
    optimization: the gray line is the full planner's frontier, the red
    line the frontier without that optimization, and the shaded region the
    frontier area lost. Percentages give the frontier area kept; the star
    is the unoptimized baseline.}
    \label{fig:pareto-ablation}
\end{figure}

\noindent\textbf{Setup.}
On both pipelines, we compare the frontier with all optimizations against the
frontier after each optimization is removed
(Figure~\ref{fig:pareto-ablation}). \emph{No Jev}
drops Jev-based implementations, so every decision is made by a
generative model or by embeddings. \emph{No batching/fusion} keeps
implementation and model choice but issues one call per tuple and never
fuses operators. \emph{No model selection} keeps one model for the whole
pipeline. Every other dimension stays available, and the baseline is
the unoptimized plan of Section~\ref{sec:optimization-effectiveness}.
We compare frontiers by their \emph{hypervolume}~\cite{zitzler1999hypervolume},
the area under the frontier, which MOBO maximizes
(Section~\ref{subsec:cost-aware-estimation}).

\noindent\textbf{Results.}
In both pipelines, batching and fusion yield the biggest gains, followed by model selection and then Jev. Without batching and fusion, plans make one call per tuple, so gains come only from weaker models or implementations, and accuracy drops when a plan must beat the most accurate one. On Misinformation, accuracy drops earliest because its fastest plans use batched LLM filtering and top-$k$. Without model selection, the pipeline must use one model; at high rates, each operator uses a weaker model and accuracy collapses. Removing Jev has the smallest effect: Stock News loses its fastest plans, and Misinformation loses some mid-accuracy plans. Jev's gains are limited because (i) at high concurrency, the serving engine amortizes per-step weight reads across calls (Eq.~\eqref{eq:latency}), making short-label generation cheap, so skipping decoding saves little, and (ii) we evaluate only one open 2B Jev model; a stronger model could route more plans to Jev.

\section{Related Work}
\label{sec:related}

\textbf{LLM-Powered Data Processing Systems and Optimizations.} A growing line of systems extends the relational model with semantic operators for processing unstructured data using LLMs~\cite{patel2025semanticoperators, shankar2025docetl, liu2024suql, lin2025docanalytics, arora2024evaporate, dspy, Lu2025VectraFlow, wei2025moar, anderson2025aryn, urban2024eleet, wang2025unify, chai2025docdb, urban2025stretto, gruber2025flockmtl}. Lotus~\cite{patel2025semanticoperators} introduces a declarative interface for semantic pipelines and optimizes individual operators via model cascades that combine a high-quality ``gold'' algorithm with a cheap proxy. Palimpzest~\cite{liu2025palimpzest} and its successor Abacus~\cite{russo2025abacus} explore cost-based optimization over models, prompts, and operator variants, formulating pipeline selection as a multi-objective optimization problem in offline, batch settings. DocETL~\cite{shankar2025docetl} applies LLM-based query rewriting to transform documents into structured representations, and MOAR~\cite{wei2025moar} extends it with multi-objective agentic rewrites searched by a multi-armed bandit. ZenDB~\cite{lin2025docanalytics} and CAESURA~\cite{urban2024caesura} target document analytics and multi-modal query planning, respectively. These systems run in batch or one-shot modes on static datasets. Some, like Abacus and MOAR, optimize quality vs. cost but pick one plan per dataset; they do not handle unbounded, evolving streams or re-planning as load and content change during persistent LLM operation.

\textbf{Stream Processing Systems.} Early systems (e.g., Aurora~\cite{abadi2003aurora}, STREAM~\cite{motwani2003query}) established the foundations of modern data-stream management, introducing core techniques for continuous queries, approximation, and adaptive resource management. These ideas were later extended by open-source engines such as Apache Flink~\cite{carbone2015flink} and Apache Storm~\cite{apache_storm}, advancing unified batch/stream execution, low-latency pipelines, and fault-tolerant processing of structured data streams.

\textbf{Adaptive and Multi-Objective Planning.} Stream systems adapt at
runtime through Eddies~\cite{avnur2000eddies}, plan
migration~\cite{zhu2004migration}, and load
shedding~\cite{tatbul2003loadshedding}; we keep pace by selecting a
cheaper plan instead of dropping tuples. Multi-objective parametric
optimization~\cite{trummer2014mpq} also picks precomputed Pareto
plans at runtime, but assumes a known cost model, and
Chameleon~\cite{jiang2018chameleon} re-profiles video analytics
configurations as content changes. In contrast, this work targets LLM-based processing of unstructured streams using continuous semantic operators and LLM-native optimizations in a dynamic planning framework. Unlike conventional systems that prioritize performance metrics, it explicitly models accuracy, enabling performance--accuracy trade-offs intrinsic to LLM-based processing.

\section{Conclusions}
\label{sec:conclusion}

Continuous Prompts (CPs) turn one-off semantic queries into persistent computation over evolving unstructured streams. With semantic windows, dynamic semantic group-by, and continuous RAG, CPs maintain evolving semantic state in the query model. A MOBO-based adaptive planner optimizes implementation, model choice, batching, and fusion to balance throughput and quality. Microbenchmarks and end-to-end streaming experiments show major speedups, effective plan selection with limited profiling, and rapid adaptation to workload and distribution shifts. CPs shift LLM data processing from one-shot prompting to continuously optimized semantic computation over unbounded streams.

New research directions follow from this work. \emph{Agentic continuous optimization} could use execution feedback to
propose and evaluate new prompts, operator compositions, and execution
strategies beyond the predefined plan space. \emph{Long-horizon semantic
pattern matching} could recognize temporal and spatial relationships
across events, such as a disruption followed by repeated delays, an
expected recovery that fails to occur, or related incidents spreading
across locations.


\bibliographystyle{IEEEtran}
\bibliography{main}

\end{document}